\documentclass[english]{emulateapj}

\usepackage{times}
\slugcomment{ApJ accepted}

\providecommand{\tabularnewline}{\\}

\newcommand{\peryr}{{\rm yr^{-1}}}

\newcommand{\LISA}{{\it LISA}}

\newcommand{\Mbh}{M_{\bullet}}

\newcommand{\Mo}{M_{\odot}}

\newcommand{\Ms}{M_{\star}}

\newcommand{\scr}{s_{\mathrm{crit}}}

\newcommand{\nbh}{n_{\bullet}}

\newcommand{\mF}{\mathcal{F}}

\shorttitle{INSPIRAL AND RELAXATION NEAR A MBH}
\shortauthors{HOPMAN \& ALEXANDER}
 
\bibpunct{}{}{}{a}{}{}

\begin{document}

\title{The orbital statistics of stellar inspiral and relaxation near a
massive black hole: characterizing gravitational wave sources}

\author{Clovis Hopman\altaffilmark{1} and Tal Alexander\altaffilmark{1,2}}

\altaffiltext{1}{Center for Astrophysics, Faculty of Physics, Weizmann
Institute of Science, POB 26, Rehovot 76100, Israel;
clovis.hopman@weizmann.ac.il, tal.alexander@weizmann.ac.il}
\altaffiltext{2}{Incumbent of the William Z. \& Eda Bess Novick career
development chair}

\begin{abstract}
We study the orbital parameters distribution of stars that are
scattered into nearly radial orbits and then spiral into a massive
black hole (MBH) due to dissipation, in particular by emission of
gravitational waves (GW). This is important for GW detection, e.g. by
the \emph{Laser Interferometer Space Antenna} (\emph{\LISA}). Signal
identification requires knowledge of the waveforms, which depend on
the orbital parameters. We use analytical and Monte Carlo methods to
analyze the interplay between GW dissipation and scattering in the
presence of a mass sink during the transition from the initial
scattering-dominated phase to the final dissipation-dominated phase of
the inspiral. Our main results are (1) Stars typically enter the
GW-emitting phase with high eccentricities. (2) The GW event rate per
galaxy is $\mathrm{few\!\times\!10^{-9}}\,\mathrm{yr^{-1}}$ for
typical central stellar cusps, almost independently of the relaxation
time or the MBH mass. (3) For intermediate mass black holes (IBHs) of
$\sim\!10^{3}\,\Mo$ such as may exist in dense stellar clusters, the
orbits are very eccentric and the inspiral is rapid, so the sources
are very short-lived.
\end{abstract}


\section{Introduction}\label{s:intro}

\setcounter{footnote}{0}

Dissipative interactions between stars and massive black holes (MBHs;
$\Mbh\!\gtrsim\!10^{6}\,\Mo$) in galactic nuclei (e.g. Gebhardt et
al. \cite{Geb00}, \cite{Geb03}), or intermediate mass black holes
(IBHs; $10^{2}\!<\!\Mbh\!\lesssim\!10^{4}\,\Mo$), which may exist
in dense stellar clusters, have been in the focus of several recent
studies.  The interest is mainly motivated by the possibility of
using the dissipated power to detect the BH, or to probe General Relativity.
Examples of such processes are tidal heating (e.g. Alexander \& Morris
\cite{AM03}; Hopman, Portegies Zwart \& Alexander \cite{HPZA04})
and gravitational wave (GW) radiation (Hils \& Bender \cite{HB95};
Sigurdsson \& Rees \cite{SR97}; Ivanov \cite{IV02}; Freitag \cite{FR01},
\cite{FR03}).

A statistical characterization of inspiral orbits is of interest in
anticipation of GW observations by the Laser Interferometer Space
Antenna (\LISA). \textit{\LISA\ } will be able to observe GW from
stars at cosmological distances during the final, highly relativistic
phase of inspiral into a $\sim\!10^{6}\,\Mo$ MBH, thereby opening
a new non-electromagnetic astronomical window. GW from inspiraling
compact objects (COs) is one of the three major targets of the \LISA\ 
mission (Barack \& Cutler \cite{BC04a}, \cite{BC04b}; Gair et al.
\cite{Gai04}), together with cosmological MBH--MBH mergers and Galactic
CO--CO mergers.

\LISA\ can detect GW emission from stars with orbital
period shorter than $P_{L}\!\sim\!10^{4}\,\mathrm{s}$.  In order for
the shortest possible period to be small enough to be detectable by
\LISA\ , the MBH has to be of moderate mass,
$\Mbh\!\lesssim\!5\!\times\!10^{6}\,\Mo$ (Sigurdsson \& Rees
\cite{SR97}). \emph{\LISA} is expected to be able to detect inspiral
into MBHs of $10^{6}\,\Mo$ to distances as far as $\gtrsim$ 1 Gpc.

The detailed time-evolution of the GW depends on the eccentricity of
the stellar orbit, and therefore probes both General Relativity and
the statistical predictions of stellar dynamics theory. Due to the low
signal to noise ratio, knowledge of the wave forms is required in
advance. For this purpose, it is necessary to estimate the orbital
characteristics of the GW-emitting stars, and in particular the
distribution function (DF) of their eccentricities (Pierro et
al. \cite{PPSLR01}; Glampedakis, Hughes \& Kennefick \cite{GHK02}), as
the wave forms are strong functions of the eccentricity (e.g. Barack
\& Cutler \cite{BC04a}; Wen \& Gair \cite{WG05}).  This study focuses
on inspiral by GW emission. However, it should be emphasized that
inspiral is a general consequence of dissipation, and the formalism
presented below can be extended in a straight-forward way to other
dissipation processes, such as tidal heating.

The \emph{prompt infall} of a star into a MBH and its destruction
have been studied extensively (\S\ref{ss:prompt}). Here we analyze
a different process, the \emph{slow inspiral} of stars (Alexander
\& Hopman \cite{AH03}). A star on a highly eccentric orbit with small
periapse $r_{p}$, repeatedly loses some energy $\Delta E$ every
periapse passage due to GW emission, and its orbit gradually decays.
At a distance $r_{0}$ from the MBH, where the orbital period is $P_{0}$,
the time-scale $t_{0}$ for completing the inspiral (i.e. decaying
to a $P\!\rightarrow\!0$ orbit) is much longer than the time-scale
$P_{0}$ needed to reach the MBH directly on a nearly radial orbit.
While the orbit decays, two-body scatterings by other stars continually
perturb it, changing its orbital angular momentum $J$ by order unity
on a timescale $t_{J}$. Because $t_{0}\!\gg\! P_{0}$, inspiraling
stars are much more susceptible to scattering than those on infall
orbits. If $t_{0}\!>\! t_{J}$, either because $r_{0}$ is large or
$r_{p}$ is large (small $\Delta E$), then the orbit will not have
time to decay and reach an observationally interesting short period.
Before that can happen, the star will either be scattered to a wider
orbit where energy dissipation is no longer efficient, or conversely,
plunge into the MBH. Inspiral is thus much rarer than direct infall.
The stellar consumption rate, and hence the properties of the stellar
distribution function (DF) at low $J$, are dominated by prompt infall,
with inspiral contributing only a small correction. This DF describes
the parent population of the inspiraling stars.

We show below that the DF of the small subset of stars on low-$J$
orbits that complete the inspiral and are GW sources is very different
from that of the parent population (Fig. \ref{f:Jlossdist}). This
results from the interplay between GW dissipation and scattering in
the presence of a mass sink during the transition ($t_{0}\!\sim\! t_{J}$)
from the initial scattering-dominated phase to the final dissipation-dominated
phase of the inspiral.

This paper is organized as follows. In \S\ref{s:lc} we recapitulate
some of the results of loss cone theory for the prompt infall, and
extend it to slow inspiral. In \S\ref{s:inspscat} and \S\ref{s:approaches}
we present a detailed analytical discussion of the main effects that
determine the rates of GW events and their statistical properties.
In \S\ref{s:approaches} we describe three different approaches for
studying the problem: by Monte Carlo simulations, by solving numerically
the 2D diffusion / dissipation equation in $E$ and $J$, and by a
simplified analytical model, which mimics the behavior of a typical
star. In \S\ref{s:results} we apply the MC simulation to a MBH in
a galactic nucleus, and an IBH in a stellar cluster. We summarize
our results in \S\ref{s:sum}.

\section{The loss-cone}

\label{s:lc}

The rate at which stars are consumed by an MBH and the effect this
has on the stellar DF near it have been studied extensively (Peebles
\cite{P72}; Frank \& Rees \cite{FR76}; Bahcall \& Wolf \cite{BW76},
\cite{BW77}; Lightman and Shapiro \cite{LS77}; Cohn \& Kulsrud \cite{CK78};
Syer \& Ulmer \cite{SU99}; Magorrian \& Tremaine \cite{MT99}; Miralda-Escud\'{e}
\& Gould \cite{MG00}; Freitag \& Benz \cite{FB01}; Alexander \& Hopman
\cite{AH03}; Wang \& Merritt \cite{WM04}; see Sigurdsson \cite{S03}
for a comparative review).  Self-consistent N-body simulations with
stellar captures were recently performed by Baumgardt, Makino, \&
Ebisuzaki (\cite{Baum04a}, \cite{Baum04b}) and by Preto, Merritt
\& Spurzem (\cite{P04}).

We begin by summarizing these results, neglecting dissipative processes.
We then extend the formalism to include dissipative processes.

\subsection{Prompt infall}

\label{ss:prompt}

The stellar orbits are defined by a specific angular momentum $J$ and
relative specific energy $\varepsilon\!=\!\psi(r)\!-\! v^{2}/2$
(hereafter {}``angular momentum'' and {}``energy''), where $\psi$ is
the relative gravitational potential, and $v$ is the velocity of a
star with respect to the MBH. A spherical mass distribution and a
nearly spherical velocity distribution are assumed.

Orbits in the Schwarzschild metric (unlike Keplerian orbits) can escape
the MBH only if their angular momentum is high enough, $J\!>\! J_{lc}(\varepsilon)$.
The phase space volume $J\!<\! J_{lc}$ is known as the {}``loss-cone''.
As argued below, stars that are scattered to low-$J$ orbits are typically
on nearly zero-energy orbits. For such orbits

\begin{equation}
J_{lc}(\varepsilon\!=\!0)=\frac{4G\Mbh}{c}\,,\label{e:Jlc}\end{equation}
The size of the loss-cone $J_{lc}$ is nearly constant over the
relevant range of $\varepsilon$.  Only during the very last in-spiral
phase the energy of the star becomes non-negligible compared to its
rest-mass, in which case the loss-cone is slightly modified (see
section {[}\ref{sub:MC}{]}). Deviations from geodetic motion due to
tidal interactions are neglected here.  This assumption is justified
for COs orbiting $\sim\!10^{6}\,\Mo$ MBHs, where the tidal radius is
much smaller than the event horizon. For main-sequence (MS) stars
where the tidal radius lies outside the event horizon, the loss-cone
is similarly defined as the minimal $J$ required to avoid tidal
disruption.

Stars that are initially on orbits with $J\!<\! J_{lc}$ will promptly
fall into the MBH on an orbital timescale. Subsequently, the infall
flow in $J$-space, $\mF(\varepsilon;J)$, is set by the rate at which
relaxation processes (here assumed to be multiple two-body scattering
events) re-populate the loss-cone orbits. 

Diffusion in $\varepsilon$-space occurs on the relaxation timescale,
$t_{r}\!\sim\!\varepsilon/\dot{\varepsilon}$, whereas diffusion in
$J$-space occurs on the angular momentum relaxation timescale,
\begin{equation} t_{J}\sim
J^{2}/\dot{\left(J^{2}\right)}\sim[J/J_{m}(\varepsilon)]^{2}t_{r}\,,\label{e:tJ}\end{equation}
where $J_{m}(\varepsilon)$ is the maximal (circular orbit) angular
momentum for specific energy $\varepsilon$.The square root dependence
of $J$ on $t_{J}$ reflects the random walk nature of the process.
Typically, $J_{lc}\!\ll\! J_{m}$. In principle, stars can enter the
loss-cone, $J\!<\! J_{lc}(\varepsilon)$ either by a decrease in $J$,
or by an increase in $\varepsilon$ (up to the last stable orbit).  In
practice, diffusion in $J$-space is much more efficient: the energy of
a star must increase by many orders of magnitude in order for it to
reach the loss-cone, which takes many relaxation times. The angular
momentum of the star, on the other hand, needs only to change by order
unity in order for the star to be captured, which happens on a much
shorter time $t_{J}\leq t_{r}$.

The ratio between $J_{lc}$ and the mean change in angular momentum
per orbit, $\Delta J$, defines two dynamical regimes of loss-cone
re-population (Lightman \& Shapiro \cite{LS77}). In the {}``Diffusive
regime'' of stars with large $\varepsilon$ (tight orbits), $\Delta J\!\ll\! J_{lc}$
and so the stars slowly diffuse in $J$-space. The loss-cone remains
nearly empty at all times since any star inside it is promptly swallowed.
At $J\!\gg\! J_{lc}$ the DF is nearly isotropic, but it falls logarithmically
to zero at $J\gtrsim J_{lc}$ (Eq. \ref{e:Nlc}). In the {}``full
loss-cone regime'' (sometimes also called the {}``pinhole'' or
{}``kick'' regime) of stars with small $\varepsilon$ (wide orbits),
$\Delta J\!\gg\! J_{lc}$ and so the stars can enter and exit the
loss-cone many times before reaching periapse. As a result, the DF
is nearly isotropic at all $J\!\gtrsim\! J_{lc}$. We argue below
that only the diffusive regime is relevant for inspiral.

The DF in the diffusive regime is described by the Fokker-Planck equation.
We follow Lightman \& Shapiro (\cite{LS77}), who neglect the small
contribution of energy diffusion to $\mF$, and write the Fokker-Planck
equation for the number density of stars $N(\varepsilon,J;t)$ as%
\footnote{We use the notation $N(x;y)$, where $x$ stands for any argument
or set of arguments (scalar of vector) and $y$ is a parameter, to
denote the stellar number density per $\mathrm{d}x$ interval at $y$.
The units of $N$ are the inverse of those of $x$. For example, $N(\varepsilon;t)\!=\!\int_{0}^{J_{m}}\mathrm{d}JN(\varepsilon,J;t)$
is the stellar number density per unit specific energy at time $t$.%
}

\begin{equation}
\frac{\partial N(\varepsilon,J;t)}{\partial
 t}=-\frac{\partial\mF(\varepsilon;J)}{\partial
 J},\label{e:FP}\end{equation} where \begin{equation}
 \mF(\varepsilon;J)=N(\varepsilon,J)\langle\Delta
 J\rangle-\frac{1}{2}\frac{\partial}{\partial
 J}N(\varepsilon,J)\langle\Delta J^{2}\rangle.\end{equation} The
 diffusion coefficients $\langle\Delta J\rangle$ and $\langle\Delta
 J^{2}\rangle$ obey the relation \begin{equation} \langle\Delta
 J^{2}\rangle=2J\langle\Delta J\rangle\qquad(J\!\ll\!
 J_{m})\,,\label{eq:Jcoeffs}\end{equation} (Lightman \& Shapiro
 \cite{LS77}; Magorrian \& Tremaine \cite{MT99}).  Much of the
 difficulty in obtaining an exact solution for the Fokker-Planck
 equation stems from the dependence of the diffusion coefficients on
 the DF; self-consistency requires solving a set of coupled equations.
 For many practical applications the diffusion coefficients are
 estimated in a non-self-consistent way, for example by assuming local
 homogeneity and isotropy (e.g. Binney \& Tremaine
 \cite{BT87}). Irrespective of its exact form, $\langle\Delta
 J^{2}\rangle$ describes a random-walk process, and is therefore
 closely related to the relaxation time.  In anticipation of the
 eventual necessity of introducing such approximations, we forgo from
 the outset the attempt to write down explicit expressions for the
 diffusion coefficients. Instead, we use them to \emph{define} the
 relaxation time $t_{r}$ as the time required to diffuse in $J^{2}$ by
 $J_{m}^{2}$,

\begin{equation}
t_{r}(\varepsilon)=\frac{J_{m}^{2}(\varepsilon)}{\langle\Delta
J^{2}\rangle}\,.\label{e:diffcoeff}\end{equation} We then treat the
relaxation time as a free parameter that characterizes the system's
typical timescale for the evolution of the DF, and whose value can be
estimated by Eqs. (\ref{e:tr_rh}, \ref{e:tr}) below.  For simplicity,
the relaxation time $t_{r}$ is assumed to be independent of angular
momentum%
\footnote{In general, if the relaxation time depends on $r$, this will
introduce some dependence of $t_r$ on $J$. We do not consider this
dependence here.}, but can generally be a function of energy.

At steady state, the stellar current $\mF(\varepsilon)$ is independent
of $J$,

\begin{equation}
\mF(\varepsilon)=\frac{1}{2}\frac{J_{m}^{2}(\varepsilon)}{t_{r}}\left[\frac{N(\varepsilon,J)}{J}-\frac{\partial N(\varepsilon,J)}{\partial J}\right]\,.\label{e:Fss}\end{equation}
Solving this equation yields\begin{equation}
N(\varepsilon,J)=-\frac{2\mF t_{r}}{J_{m}(\varepsilon)^{2}}J\ln J+CJ\,.\label{e:solnoBC}\end{equation}
 The integration constants $C$ and $\mF$ that are determined by
the boundary conditions $N(\varepsilon,J_{lc})\!=\!0$ and $N(\varepsilon,J_{m})\!=\! N_{\mathrm{iso}}(\varepsilon,J_{m})$.
The isotropic DF is separable in $\varepsilon$ and $J$,

\begin{equation}
N_{\mathrm{iso}}(\varepsilon,J)\mathrm{d}\varepsilon\mathrm{d}J=\frac{2N_{\mathrm{iso}}(\varepsilon)J}{J_{m}^{2}(\varepsilon)}\mathrm{d}\varepsilon\mathrm{d}J\,.\label{e:Niso}\end{equation}
 Applying these boundary conditions\footnote{Adiabatic MBH growth may
 lead to some anisotropy (Quinlan, Hernquist \& Sigurdsson
 \cite{QHS95}.). Here we assume that far from the loss-cone the DF will
 be isotropic} to equation (\ref{e:solnoBC}), the DF is given by

\begin{equation}
N(\varepsilon,J)=\frac{2N_{\mathrm{iso}}(\varepsilon)J}{J_{m}^{2}(\varepsilon)}\frac{\ln(J/J_{lc})}{\ln(J_{m}/J_{lc})},\label{e:Nlc}\end{equation}
 and the stellar current into the MBH per energy interval is

\begin{equation}
\mF(\varepsilon)=\frac{N_{\mathrm{iso}}(\varepsilon)}{\ln(J_{m}/J_{lc})t_{r}(\varepsilon)}\,.\label{e:Flow}\end{equation}
 Note that the capture rate in the diffusive regime depends only logarithmically
on the size of the loss cone. 

The prompt infall rate $\Gamma_{p}$ in the diffusive regime is then
given by

\begin{equation}
\Gamma_{p}=\int_{\varepsilon_{p}}^{\infty}\frac{\mathrm{d}\varepsilon N_{\mathrm{iso}}(\varepsilon)}{\ln(J_{m}/J_{lc})t_{r}(\varepsilon)}\,,\label{e:Gp}\end{equation}
 where the energy $\varepsilon_{p}$ separates the diffusive and full
loss-cone regimes. A star samples all angular momenta $J_{lc}\!<\! J\!<\! J_{m}$
in a relaxation time, and it is promptly captured once $J\!<\! J_{lc}$.
The total rate is therefore of order $\Gamma_{p}\!\sim\! N_{\mathrm{iso}}(<\! a_{p})/t_{r}(a_{p})$,
where $a_{p}$ is the typical radius associated with orbits of energy
$\varepsilon_{p}$ ($a_{p}$ is the semi-major axis for Keplerian
orbits, see \S\ref{ss:KeplerCusp}) and $N_{\mathrm{iso}}(<\! a_{p})$
is the number of stars within $a_{p}$. The rate is logarithmically
suppressed because of the diluted occupation of phase space near the
loss cone.

\subsection{Keplerian orbits in a power-law cusp}

\label{ss:KeplerCusp}

The MBH dominates the stellar potential within the radius of influence,

\begin{equation}
r_{h}=\frac{G\Mbh}{\sigma^{2}},\label{e:rh}\end{equation} where
 $\sigma^{2}$ is the 1D stellar velocity dispersion far from the
 MBH. The mass enclosed within $r_{h}$ is roughly equal to $\Mbh$.
 Various formation scenarios predict that the spatial stellar number
 density at $r\!<\! r_{h}$ should be approximately a power law (e.g.
 Bahcall \& Wolf \cite{BW76}; Young \cite{You80}) \begin{equation}
 n_{\star}(r)=\frac{(3/2-p)N_{h}}{4\pi
 r_{h}^{3}}\left(\frac{r}{r_{h}}\right)^{-3/2-p}\,,\label{e:nplaw}\end{equation}
 where $N_{h}$ is the number of stars inside $r_{h}$. This corresponds
 to an energy distribution
 $N(\varepsilon)\mathrm{d}\varepsilon\!\propto\!\varepsilon^{p-5/2}\mathrm{d}\varepsilon$.
 A stellar cusp with $p\!\sim\!0$ has been observed in the Galactic
 Center (Alexander \cite{TA99}; Genzel et al. \cite{Genzelea03}).  For
 a single mass population, it was shown by analytical considerations
 that $p=1/4$ (Bahcall \& Wolf \cite{BW76}). This has been confirmed
 recently by N-body simulations (Baumgardt et al. \cite{Baum04a};
 Preto et al. \cite{P04}).

Mass segregation drives the heavy stars in the population to the center.
The radial distribution of the different mass components can be then
approximated as average power-laws, steeper ($p\!>\!0$) for the heavier
masses and flatter ($p\!\lesssim\!0$) for the lower masses (Bahcall \&
Wolf \cite{BW77}; Baumgardt et al. \cite{Baum04b}).

Typically, the diffusive regime is within the radius of influence.
We therefore assume from this point on that the stars move on Keplerian
orbits ($\psi\!=\! G\Mbh/r$) in a power-law density cusp. The stellar
orbits are characterized by a semi-major axis $a$, eccentricity $e$,
periapse $r_{p}$ and period $P$,

\begin{eqnarray}
a\!=\frac{G\Mbh}{2\varepsilon}\,,\qquad\:\:\:\:\:\: &  & e^{2}\!=\!1-\frac{J^{2}}{G\Mbh a}\,,\nonumber \\
r_{p}\!=\! a(1-e)\,,\qquad &  & P\!=\!\frac{2\pi a^{3/2}}{\sqrt{G\Mbh}}\,.\label{eq:Kepler}\end{eqnarray}
 During most of the inspiral $1\!-\! e\!\ll\!1$, and the periapse
can be approximated by $r_{p}\!\approx\! J^{2}/2G\Mbh$. This remains
valid until the last phases of the inspiral. 

The prompt infall rate (Eq. \ref{e:Gp}) can be expressed in terms
of the maximal semi-major axis, \begin{equation}
\Gamma_{p}=\int_{0}^{a_{p}}\frac{\mathrm{d}aN_{\mathrm{iso}}(a)}{\ln(J_{m}/J_{lc})t_{r}(a)}\,.\label{eq:prompt}\end{equation}

We will assume Keplerian orbits throughout most of this paper except
for section (\ref{ss:MC}), where we employ the general relativistic
potential of the MBH.

\subsection{Slow inspiral}

\label{ss:inspiral}

The derivations of the conditions necessary for slow inspiral and
of the inspiral rate follow closely those of prompt infall, but with
two important differences. (1) The time to complete the inspiral is
not the infall time $P_{0}$, but rather $t_{0}\!\gg\! P_{0}$. (2)
There is no contribution from the full loss-cone regime, where stars
are scattered multiple times each orbit. This is because inspiral
in this regime would require that the \emph{very same} star that was
initially deflected into an eccentric orbit, be re-scattered back
into it multiple times. The probability for this happening is effectively
zero%
\footnote{This is to be contrasted with prompt infall from the full loss-cone
regime, where a star can reach the MBH by being scattered \emph{once}
into the loss-cone just before crossing $a_{p}$ toward the MBH. %
}.

In analogy to the radial scale $a_{p}$ of prompt infall, which delimits
the volume where stars can avoid scattering for a time $P_{0}\!<\! t_{J}$,
and thus maintain their infall orbit until they reach the MBH, the
inspiral criterion $t_{0}\!<\! t_{J}$ defines a critical radius $a_{c}\!\ll\! a_{p}$
(or equivalently, a critical energy $\varepsilon_{c}\!\gg\!\varepsilon_{p}$).
Stars starting the inspiral from orbits with $a_{0}\!<\! a_{c}$ ($\varepsilon_{0}\!>\!\varepsilon_{c}$)
will complete it with high probability, whereas stars starting with
$a_{0}\!>\! a_{c}$ ($\varepsilon_{0}\!<\!\varepsilon_{c}$), will
sample all $J$ values before they spiral in significantly, \emph{regardless
of $J_{0}$} and ultimately either (1) fall in the MBH, (2) diffuse
in energy to much wider orbits or (3) into the much tighter orbits
of the diffusive regime. Since we assume a steady state DF, outcomes
(2) and (3) represent a trivial, DF-preserving redistribution of stars
in phase space, which does not affect the statistical properties of
the system. Whether stars spiral in or fall in depends, statistically,
only on $a_{0}$ ($\varepsilon_{0}$). We use below Monte Carlo simulations
(\S\ref{ss:MC}, Fig. \ref{f:GWfrac}) to estimate the inspiral probability
function, $S(a_{0})$, which describes the probability of completing
inspiral when starting from an orbit with semi-major axis $a_{0}$
($S\!\rightarrow\!1$ for $a_{0}\!\ll\! a_{c}$, $S\!\rightarrow\!0$
for $a_{0}\!\gg\! a_{c}$). The inspiral rate for stars of type $s$
with number fraction $f_{s}$ is then 

\begin{equation}
\Gamma_{i}=f_{s}\int_{0}^{\infty}\!\frac{\mathrm{d}aN(a)S(a)}{\ln(J_{m}/J_{lc})t_{r}(a)}\simeq f_{s}\int_{0}^{a_{c}}\!\frac{\mathrm{d}aN(a)}{\ln(J_{m}/J_{lc})t_{r}(a)}\,,\label{e:SU}\end{equation}
where roughly $S(a_{c})\!\sim\!0.5$.

\section{Parameter dependence of the inspiral rate}

\label{s:inspscat}

In this section we derive some analytical results for the inspiral
rate. In order to keep the arguments transparent, we neglect relativistic
deviations from Keplerian motion. Relativistic orbits are discussed
in section (\ref{ss:MC}).

Consider a star of mass $\Ms$ orbiting a MBH of mass $\Mbh$ on a
bound Keplerian orbit with semi-major axis $a$ and angular momentum
$J$. When the star arrives at periapse, it loses some orbital energy
$\Delta E$ by GW emission. As a result, the orbit shrinks and its
energy increases. For highly eccentric orbits the periapse of the
star is approximately constant during inspiral in absence of scattering.
We define the inspiral time $t_{0}$ as the time it takes the initial
energy $\varepsilon_{0}$ to grow formally to infinity. If the energy
loss per orbit is constant, then for $e\rightarrow1$

\begin{equation}
t_{0}=\int_{\varepsilon_{0}}^{\infty}\frac{\mathrm{d}\varepsilon}{(\mathrm{d}\varepsilon/\mathrm{d}t)}\approx\frac{1}{\Delta E}\int_{\varepsilon_{0}}^{\infty}\mathrm{d}\varepsilon P(\varepsilon)=\frac{2\varepsilon_{0}P_{0}}{\Delta E},\end{equation}
 or \begin{equation}
t_{0}(r_{p},a)=\frac{2\pi\sqrt{GM_{\bullet}a}}{\Delta E}.\label{e:t0}\end{equation}

For GW, $\Delta E$ is given by (Peters \cite{Pe64}) 

\begin{equation}
\Delta E_{\mathrm{GW}}=\frac{8\pi}{5\sqrt{2}}f(e)\frac{\Ms c^{2}}{\Mbh}\left(\frac{r_{p}}{r_{S}}\right)^{-7/2}\,,\label{e:DEgw}\end{equation}
where\begin{equation}
f(e)=\frac{1+\frac{73}{24}e^{2}+\frac{37}{96}e^{4}}{(1+e)^{7/2}}\,,\label{e:f(e)}\end{equation}
and $r_S=2G\Mbh/c^2$ is the Schwarzschild radius. 

During all but the last stages of the inspiral $e\!\sim\!1$, in which
case $r_{p}/r_{S}=4(J/J_{lc})^{2}$ and

\begin{equation}
\Delta E_{\mathrm{GW}}=E_{1}(J/J_{lc})^{-7}\,,\qquad E_{1}\equiv\frac{85\pi}{3\!\times\!2^{13}}\frac{\Ms c^{2}}{\Mbh}\,.\label{e:DEgwJ}\end{equation}
 Gravity waves also carry angular momentum, \begin{equation}
\Delta J_{\mathrm{GW}}=-\frac{16\pi}{5}g(e)\frac{G\Ms}{c}\left(\frac{r_{p}}{r_{S}}\right)^{-2}\,,\label{e:DJgw}\end{equation}
\begin{equation}
g(e)=\frac{1+\frac{7}{8}e^{2}}{(1+e)^{2}}\,.\label{e:g(e)}\end{equation}
Generally, the change in $J$ in the course of inspiral is dominated
by two-body scattering, and $\Delta J_{\mathrm{GW}}$ can be neglected
until $a$ becomes very small.

It is convenient to refer the timescales in the system to the relaxation
time at the MBH radius of influence,

\begin{equation}
t_{h}=A_{p}\left(\frac{\Mbh}{\Ms}\right)^{2}\frac{P(r_{h})}{N_{h}\mathrm{log}\Lambda_{1}}\,,\label{e:tr_rh}\end{equation}
where $\Lambda_{1}\!=\!\Mbh/\Ms$ $(r_{S}/r_{h})^{1/4}$ (Miralda-Escud\'{e}
\& Gould \cite{MG00}), and $A_{p}\!\simeq\!0.2$ for $p\!=\!0$
(Alexander \& Hopman \cite{AH03}). The relaxation time at any radius
is then

\begin{equation}
t_{r}(a)=t_{h}\left(\frac{a}{r_{h}}\right)^{p}\,,\label{e:tr}\end{equation}
 where we associate the typical relaxation time on an orbit with that
 at its semi-major axis. This is a good approximation, since
 theoretical arguments (Bahcall \& Wolf \cite{BW76}; \cite{BW77}) and
 simulations (Freitag \& Benz \cite{Fre02}; Baumgardt et
 al. \cite{Baum04a} \cite{Baum04b}; Preto et al. \cite{P04}) indicate
 that $0\lesssim p\lesssim0.25$, and so $t_{r}$ is roughly independent
 of radius. The angular momentum relaxation time is

\begin{equation}
t_{J}=\left[\frac{J}{J_{m}(a)}\right]^{2}\left(\frac{a}{r_{h}}\right)^{p}t_{h}\,.\label{e:tp}\end{equation}

Dissipational inspiral takes place in the presence of two-body scattering.
When $t_{0}\!\sim\! t_{J}$, both effects have to be taken into account.
It is useful to parametrize the relative importance of dissipation
and scattering by the dimensionless quantity

\begin{equation}
s(J,a)\equiv\frac{t_{0}(J,a)}{t_{J}(J,a)}=\left(\frac{a}{d_{c}}\right)^{3/2}\left(\frac{a}{r_{h}}\right)^{-p}\left(\frac{J}{J_{lc}}\right)^{5},\label{e:s}\end{equation}
 where we introduce the ($p$-independent) length scale\begin{equation}
d_{c}\equiv\left(\frac{8\sqrt{G\Mbh}E_{1}t_{h}}{\pi c^{2}}\right)^{2/3}\,,\label{e:D}\end{equation}
 which is of the same order as $a_{c}$ (Eq. \ref{e:ac}) and is $\ll\! r_{h}$.
We define some critical value $s_{\mathrm{crit}}\!\ll\!1$ such that
the inspiral is so rapid that the orbit is effectively decoupled from
the perturbations.

The three phases of inspiral can be classified by the value of $s$.
In the {}``scattering phase'' the star is far from the region in
phase space where GW emission is efficient and $s\!\gg\!1$. With
time it may scatter to a lower-$J$ orbit, enter the {}``transition
phase'', where $s\!\sim\!1$, and start to spiral in. If it is not
scattered into the MBH or to a wide orbit, it will eventually reach
the stage where $s\!<\! s_{\mathrm{crit}}$. It will then enter the
{}``dissipation phase'' where it spirals-in deterministically according
to equations (\ref{e:DEgw}--\ref{e:g(e)}). Note that eventually
the $e\!\rightarrow\!1$ approximation is no longer valid. 

Here we are mainly interested in understanding how the interplay between
two-body scattering and energy dissipation in the first two phases
sets the initial conditions for the GW emission in the final phase.
It should be emphasized that the onset of the dissipation phase does
not necessarily coincide with the emission of \emph{detectable} GW.
For example, while a star is well into the dissipation phase by the
time $s\!<\! s_{\mathrm{crit}}\!\sim\!10^{-3}$, the orbit has still
to decay substantially before the GW frequency becomes high enough
to be detected by \LISA. 

We derive an analytical order of magnitude estimate for the critical
semi-major axis $a_{c}$ by associating it with $s\!=\!1$ orbits
in the transition phase. Since $s$ falls steeply with $J$, we set
$J\!=\! J_{lc}$ and solve $ $$s(J_{lc},a_{c})\!=\!1$ for $a_{c}$,
obtaining 

\begin{equation}
\frac{a_{c}}{r_{h}}=\left(\frac{d_{c}}{r_{h}}\right)^{3/(3-2p)}.\label{e:ac}\end{equation}
 The MC simulations below (\S\ref{ss:MC}) confirm that this analytical
estimate corresponds within a factor of order unity to the semi-major
axis where the inspiral probability $S(a_{c})\!\sim\!0.5$, for a
wide range of masses (see table {[}\ref{t:stars}{]}). Expression
(\ref{e:SU}) for the rate can then be approximated by

\begin{equation}
\Gamma_{i}\sim\frac{f_{s}N_{h}}{t_{h}\ln[J_{m}(a_{c})/J_{lc}]}\left(\frac{a_{c}}{r_{h}}\right)^{3/2-2p},\label{e:Gammai}\end{equation}
 where $N_{h}$ is the number of stars within $r_{h}$. 

The dependence of the inspiral rate on $p$ (at fixed $d_{c}/r_{h}$
and neglecting the logarithmic terms) can be examined by writing $\Gamma_{i}\!\simeq\! g(p)f_{s}(N_{h}/t_{h})$,
where

\begin{equation}
g(p)=\left(\frac{d_{c}}{r_{h}}\right)^{\left.3(3-4p)\right/2(3-2p)}.\label{e:ggamma}\end{equation}
The pre-factor $g(p)$ grows with $p$ over the relevant range (see
also Ivanov \cite{IV02}). This reflects the fact that the inspiral
rate is determined by the number of stars within $a_{c}$, rather
than the total number of stars within $r_{h}$ . The concentration
of the cusp increases with $p$, so that there are more stars within
$a_{c}$. This result suggests qualitatively that in a mass-segregated
population, the heavier stars (higher $p$) will have an enhanced
GW event rate compared to the light stars (lower $p$).

>From equations (\ref{e:D}--\ref{e:ggamma}) it follows that \begin{equation}
\Gamma_{i}\propto t_{h}^{-2p/(3-2p)}.\end{equation}
Since $p\!\sim\!0$ for typical stellar cusps around MBHs, this means
that the inspiral rate is nearly independent of the relaxation time
for such cusps. This counter-intuitive result reflects the near balance
between two competing effects. When scattering is more efficient,
stars are supplied to inspiral orbits at a higher rate, but are also
scattered off them prematurely at a higher rate, so the volume of
the diffusive regime, which contributes stars to the inspiral ($\sim\! a_{c}^{3}$),
decreases. This is in contrast to the prompt disruption rate $\Gamma_{p}$,
which increases as the relaxation time becomes shorter%
\footnote{The prompt disruption rate is $\Gamma_{p}\!\sim\!\left.N(<\! a_{p})(r_{S}/a_{p})\right/P(a_{p})$
(Syer \& Ulmer \cite{SU99}, Eq, 10). For prompt disruption $a_{p}\!\sim\! t_{h}^{2/(5-2p)}$
(Alexander \& Hopman \cite{AH03}), and so the rate scales as $\Gamma_{p}\!\sim\! t_{h}^{-(2+2p)/(5-2p)}$
.%
}.  It then follows that enhanced scattering, such as by massive perturbers
(e.g. clusters, giant molecular clouds; Zhao, Haehnelt \& Rees \cite{ZHR02})
increases the prompt disruption rate, but will not enhance the rate
of inspiral events.

The dependence of the GW event rate on the mass of the MBH can be
estimated from Eq. (\ref{e:Gammai}) and the empirical $M_{\bullet}$--$\sigma$
relation \begin{equation}
M_{\bullet}=1.3\times10^{8}M_{\odot}\left(\frac{\sigma}{200\,\mathrm{km\, s^{-1}}}\right)^{\beta},\label{e:Msigma}\end{equation}
 where $\beta\!\sim\!4$ (Ferrarese \& Merritt \cite{FM00}; Gebhardt
et al. \cite{Geb00}; Tremaine et al. \cite{Tr02}). Note that the
$\Mbh$--$\sigma$ relation implies that the stellar number density
at the radius of influence, $n_{h}$ is larger for lighter MBHs: for
example, for $\beta\!=\!4$, $n_{h}\!\propto\! N_{h}r_{h}^{-3}\!\propto\!\Mbh^{-1/2}$,
where we assumed that $N_{h}\!\propto\!\Mbh$; the consequences of
this for the dependence of the rate for prompt tidal disruptions on
the MBH mass were discussed by Wang \& Merritt (\cite{WM04}).

The GW event rate depends on $\Mbh$ as 

\begin{equation}\label{eq:massdependence}
\Gamma_{i}\propto\Mbh^{3/\beta-1}\,.\label{e:massdependence}\end{equation}
This dependence is weak, e.g. $\Gamma_{i}\!\propto\!\Mbh^{-1/4}$for
$\beta=4$. Thus, the rate becomes \textit{higher} for lower mass
MBHs. If $\Mbh\sim10^{3}\,\Mo$ IBHs indeed exist in stellar clusters
and the $M_{\bullet}$--$\sigma$ relation can be extrapolated to
these masses, then they may be more likely to capture stars than MBHs.
This, however, does not necessarily translate into more GW sources.
The strain $h$ of GW decreases with the mass of the MBH, so that
these sources have to be closer by in order to observe their GW emission.
Another restriction is that for IBHs the tidal force is so strong
that white dwarfs are tidally disrupted well outside the event horizon,
which precludes them from being \emph{\LISA} sources. These issues
are further discussed in \S\ref{s:results}.

\section{Orbital evolution with dissipation and scattering}

\label{s:approaches} We present three different methods for analyzing
inspiral in the presence of scattering. The first approach is based
on Monte Carlo (MC) simulations, which follow a star on a relativistic
orbit, described by $\varepsilon$ and $J$, and add small perturbations
to simulate energy dissipation and random two-body scattering. The
second approach consists of direct numerical integration of the time
dependent diffusion-dissipation equation. The third approach is a
heuristic semi-analytical effective model that can describe the {}``effective''
trajectory of a star through phase space, as well as the statistical
properties of an ensemble of such trajectories.

The three approaches are complementary. The MC simulations allow a
direct realization of the micro-physics of the system, since they
follow the perturbed orbits of individual stars. They also offer much
flexibility in setting the initial conditions of the numerical experiments,
but the results are subject to statistical noise and are not easy
to generalize. The diffusion-dissipation equation on the other hand
deals directly with the DF, and allows an analytical formulation of
the problem in terms of partial differential equations, which are
solved numerically. However, computational limitations do not allow
covering as large a dynamical range as in the MC simulations. Finally,
the heuristic effective model has the advantage of its intuitive directness
and relative simplicity of use. We find that all three methods give
the same results for the same underlying assumptions (Fig. \ref{f:Jlossdist}).
This inspires confidence in the robustness of the analysis.

To compare the three methods, we stop the simulation at the point
where $s=s_{{\rm crit}}=10^{-3}$, and we plot the DF of the angular
momenta of the stars. From that point on the stars are effectively
decoupled from the cluster and can spiral in undisturbed. We then
use the MC method, which can be easily extended to follow the stars
in the dissipation phase (section \ref{s:results}), to find the DF
of the eccentricities of stars which enter the \emph{\LISA} band.

\subsection{Monte Carlo simulations\label{sub:MC}}

\label{ss:MC}

The MC simulations generally follow the scheme used by Hils \& Bender
(\cite{HB95}) to study the event rate of GW. The star starts on an
initial orbit with $(\varepsilon_{0},J_{0})$ such that dissipation by
GW emission is negligible. The initial value of $J$ is not of
importance as the angular momentum is quickly randomized in the first
few steps of the simulation. Every orbital period $P(\varepsilon)$,
the energy and angular momentum are modified by $\delta\varepsilon$
and $\delta J$, and the orbital period and periapse are recalculated
(this diffusion approach is justified as long as $P\!\ll\! t_{J}$).
The simulation stops when the star decays to an orbit with $s\!<\!
s_{\mathrm{crit}}\!=\!10^{-3}$ or when $J\!<\! J_{lc}$ and it falls in
the MBH (escape to a less bound orbit is not an option here since
energy relaxation is neglected and only dissipation is considered; see
\S\ref{ss:inspiral}).

When stars reach high eccentricities as a result of scattering, the
periapse approaches the Schwarzschild radius to the point where the
Newtonian approximation breaks down.The MC simulations take this into
account by integrating the orbits in the relativistic potential of
a Schwarzschild MBH. The periapse of the star moving in a Schwarzschild
spacetime is related to its angular momentum by one of the three roots
of the equation\begin{equation}
E_{{\rm GR}}^{2}=\left(c^{2}-\frac{2GM_{\bullet}}{r}\right)\left(c^{2}+\frac{J^{2}}{r^{2}}\right)\equiv V_{{\rm GR}}(r,J).\label{eq:GRpot}\end{equation}
The term on left hand side of this equation is the squared specific
relativistic energy of the star (including its rest mass), while the
right hand side is the effective GR potential. A star on a bound orbit
is not captured by the MBH as long as equation (\ref{eq:GRpot}) has
three real roots for $r$. The smallest root is irrelevant for our
purposes. The intermediate root (the turning-point) is the periapse
$r_{p}$ of the orbit, and the largest root is the apo-apse $r_{a}$.
The semi-major axis $a$ and eccentricity $e$ are defined by $a=(r_{p}+r_{a})/2$
and $e=1-r_{p}/a$ (see Eq. \ref{eq:Kepler}). For bound orbits the
loss cone $J_{lc}$ is a very weak function of energy and is well
approximated by Eq. (\ref{e:Jlc}).

For given values of $E$ and $J$, the GR periapse is smaller than
the Newtonian one, and therefore the orbits can be more eccentric,
and the dissipation can be stronger than implied by the Newtonian
approximation. The Keplerian relations between energy, angular momentum
and the orbital parameters (Eqs \ref{eq:Kepler}) are replaced by 

\begin{equation}
E_{{\rm GR}}^{2}=\frac{(q-2-2e)(q-2+2e)}{q(q-3-e^{2})};\label{eq:Egr}\end{equation}
\begin{equation}
J^{2}=\frac{q^{2}}{q-3-e^{2}}\left(\frac{GM_{\bullet}}{c}\right)^{2},\label{eq:Jgr}\end{equation}
where $q=2(1-e^{2})a/r_{S}$ (Cutler, Kennefick \& Poisson
\cite{CKP94}).  The condition for a non-plunging orbit can also be
expressed in terms of the eccentricity and the semi-major axis,
$2(1-e^{2})a=(6+2e)r_{S}$ (Cutler et al. \cite{CKP94}). This
corresponds to a maximal eccentricity for a star on a non-plunging
orbit\begin{equation} e_{{\rm
max}}(a)=-\frac{r_{S}}{2a}+\sqrt{(r_{S}/2a)^{2}-3r_{S}/a+1},\label{e:eL}\end{equation}
which increases with $a$. If
$a_{L}=(P_{L}^{2}/4\pi^{2}GM_{\bullet})^{1/3}$ is the maximal
semi-major axis a star may have in order to be detectable by \LISA\ ,
the maximal eccentricity of a star detectable by \LISA\ is
$e_{\max}(a_{L})$.

The step in $J$-space per orbit is the sum of three terms \begin{equation}
\delta J(\varepsilon,J)=\Delta_{1}J_{{\rm scat}}(\varepsilon,J)+\chi\Delta_{2}J_{\mathrm{scat}}(\varepsilon,J)-\Delta J_{\mathrm{GW}}(\varepsilon,J)\,.\label{e:dJ}\end{equation}
 The first and second terms represent two-body scattering (Eqs. \ref{e:tJ},\ref{eq:Jcoeffs},\ref{e:diffcoeff}),
with%
\footnote{The drift term $\Delta_{1}J_{{\rm scat}}$ represents a bias to scatter
\emph{away} from the MBH due to the 2D character of the direction
of the velocity vector $\mathbf{v}$. This can be expressed geometrically
by considering a circle of radius $\Delta v_{\mathrm{scat}}$ centered
on $\mathbf{v}$ ($\mathbf{v}\!+\!\Delta\mathbf{v}{}_{{\rm scat}}$is
the change per orbit due to scattering). This circle is intersected
by a second circle of radius $v$ that passes through $\mathbf{v}$
and is centered on the radius vector to the MBH. The section of the
first circle that leads away from the MBH is slightly larger than
the section leading toward it. %
} $\Delta_{1}J_{{\rm scat}}\!=\!\left\langle \Delta J\right\rangle P\!=\! J_{m}^{2}P(\varepsilon)/(2t_{r}J)$
and $\Delta_{2}J_{\mathrm{scat}}\!=\!\sqrt{\left\langle \Delta J^{2}\right\rangle P}\!=\!\sqrt{P(\varepsilon)/t_{r}}J_{m}(\varepsilon)\!=\!\sqrt{P(\varepsilon)/t_{J}}J$.
The random variable $\chi$ takes the values $\pm1$ with equal probabilities.
The third term is the deterministic angular momentum loss due to GW
emission (Eq. \ref{e:DJgw}).The energy step per orbit is deterministic
(diffusion in energy is neglected)

\begin{equation}
\delta E(\varepsilon,J)=\Delta
E_{\mathrm{GW}}(\varepsilon,J)\,.\label{e:dE}\end{equation} In
order to increase the speed of the simulation we use an adaptive
time-step. We checked for some cases that taking a smaller time-step
does not affect the results.

The DF of inspiraling stars is generated by running many simulations
(typically $3\!\times\!10^{4}$) \texttt{\textbf{}}of stellar
trajectories through phase space with the same initial value for $J_0$
for given initial semi-major axis $a_0$, but with different random
perturbations. We verified that the initial value of $J_0$ is
irrelevant, as long as $J_0\gg J_{lc}$. We record the fraction of
stars that avoid falling in the MBH, $S(a_{0})$, and the value of $J$
at $s_{\mathrm{crit}}$ for those stars that reach the
dissipation-dominated phase, thereby obtaining the DF
$W(J;a_{0})$. This is repeated for a range of $a_{0}$ values. The
integrated DF over all cusp stars, $W(J)$, is then obtained by taking
the average of all DFs, weighted by $N(a_{0})S(a_{0})\mathrm{d}a_{0}$
(cf Eq. \ref{e:SU}).

\subsection{Diffusion equation with GW dissipation}

\label{ss:DiffDiss} 

The DF of the inspiraling stars at the onset of the dissipation dominated
phase can be obtained directly by solving the diffusion-dissipation
equation. This approach was taken by Ivanov (\cite{IV02}), who included
a GW dissipation term and obtained analytic expressions for the GW
event rate for $J\!\gg\! J_{lc}$ under various simplifying assumptions.
Here we are interested DF of stars very near the loss-cone, and so
we integrate the diffusion-dissipation equation numerically with two
simplifying assumptions. (1) We neglect the drift term in the diffusion-dissipation
equation. This can be justified by noting that the drift grows linearly
with time, $\delta J_{1}(t)=J_{m}^{2}t/(2t_{r}J)$, while the change
due to the random walk grows as $\delta J_{2}(t)=(t/t_{r})^{1/2}J_{m}$.
For a star with initial angular momentum $J$ the drift becomes important
only for $t\!>\! t_{{\rm drift}}=4(J/J_{m})^{2}t_{r}\!=\!4t_{J}$.
Since we are interested in the timescales $t\!\sim\! t_{0}\le\! t_{J}$,
the drift is only a small correction. (2) We assume Keplerian orbits.
These approximations are validated by the very good agreement with
the MC simulation results (section \ref{ss:compare}).

With these two assumptions, the diffusion-dissipation is (cf Eqs.
\ref{e:FP}, \ref{e:Fss})

\begin{eqnarray}
\frac{\partial N(\varepsilon,J;t)}{\partial t} & = & \frac{J_{m}^{2}(\varepsilon)}{2t_{r}}\frac{\partial^{2}}{\partial J^{2}}N(\varepsilon,J;t)\nonumber \\
 & + & \frac{\partial}{\partial\varepsilon}\left[\dot{E}_{\mathrm{GW}}(J)N(\varepsilon,J;t)\right],\label{e:diffdiss}\end{eqnarray}
 where $\dot{E}_{\mathrm{GW}}=\Delta E_{\mathrm{GW}}/P$ is the rate
at which energy is lost to GW. The first term accounts for diffusion
in $J$-space and the second represents the energy dissipated by GW
emission. As with the MC simulations, the diffusion coefficient $\left\langle \Delta J^{2}\right\rangle \!=\! J_{m}^{2}/t_{r}$
is an input parameter rather than resulting from a self-consistent
calculation.

The initial conditions consist of an isotropic cusp DF, $N(\varepsilon,J)\propto\varepsilon^{p-5/2}J$
and the boundary condition that the DF vanish on the loci $s\!=\!\scr\!=\!10^{-3}$
and $J\!=\! J_{lc}$ in the ($\varepsilon$,$J$)-grid. The initial
DF is evolved in time over the $(\varepsilon,J)$-grid until $t\!\sim\! t_{J}$,
when a relaxed steady state is achieved. The integration is done using
a Forward Time Centered Space (FTCS) representation of the diffusion
term, and an 'upwind' differential scheme for the dissipation term
(see e.g. Press et al. \cite{PTVF77}). After each time step (which
are chosen small enough to obey the Courant condition) the DF is re-normalized
so that captured stars are replenished.

After relaxation, the DF is non-zero up to the boundary $s\!=\!\scr$,
as stars are redistributed in phase space by diffusion at large $J$
and by dissipation at small $J$. The DF at $s_{\mathrm{crit}}$ is
then extracted to construct the angular momentum DF, $W(J)$. Although
the stars continue their trajectory in phase space beyond $s_{\mathrm{crit}}$
all the way to the last stable orbit, this does not change $W(J)$
because the rapid energy dissipation does not allow much $J$-diffusion.

\subsection{Analytical model of effective orbit }

\label{s:model}

\begin{figure}

\plotone{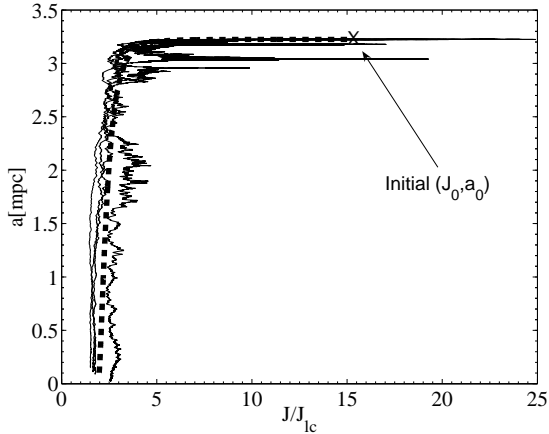}

\figcaption[FileName]{\label{f:Ja}Examples of tracks of stars from the MC simulation (solid
lines), and a track from the effective model (dashed line) with $\xi=1$.
The effective track represents the stochastic tracks well. The tracks
extend to the point where $s\!=\!10^{-3}$. The relevant parameters
are given in the first line of table (\ref{t:stars}), i.e. inspiral
of a white dwarf into a MBH.}
\end{figure}

Stars follow complicated stochastic tracks in $(J,a)$ space; due
to scattering they move back and forth in $J$, while they always
drift to smaller $a$ due to GW dissipation (energy diffusion by scattering
is neglected). The drift rate depends strongly on $J$. Figure (\ref{f:Ja})
shows some examples of stellar tracks in phase space, taken from the
MC simulations.

In this section we propose a heuristic analytical model, which captures
some essential results of the MC simulations, while providing a more
intuitive understanding.

In this model we define the equations of motion of an {}``effective
track'' of a star, which is deterministic and can be solved analytically.
With this method we follow stars that start at some given initial
semi-major axis $a_{0}$ and angular momentum $J_{0}$, and follow
the star during its first two stages of inspiral, i.e., until the
moment that $s\!<\! s_{\mathrm{crit}}$.

The DF of stars which reach $s\!=\! s_{\mathrm{crit}}$ is determined
by the evolution of the orbital parameters ($J,a$) in the region
where energy dissipation is efficient. Because of the strong dependence
of the energy $\Delta E$ which is dissipated per orbit on angular
momentum (or, equivalently, periapse), this is a small region $\Delta J$
in $J$-space, the size of which is of order of the loss-cone, $\Delta J\!\sim\! J_{lc}$.
The time $\Delta t$ it typically takes a star with semi-major axis
$a_{0}$ performing a random walk in $J$ to cross this region is
$\Delta t\!=\![\Delta J/J_{m}(a_{0})]^{2}t_{h}$ ($t_{r}\!=\! t_{h}$
assumed). This can be used to introduce an effective $J$-{}``velocity''
$\Delta J/\Delta t\!=\!-J_{m}^{2}(a_{0})/J_{lc}t_{h}$.

The effective velocity of the semi-major axis is given by

\begin{equation}
\frac{\mathrm{d}a(J,a)}{\mathrm{d}t}=\frac{da}{dE}\dot{E}=-\frac{2a^{2}}{G\Mbh}\frac{\Delta E_{\mathrm{GW}}(J)}{P(a)}.\label{e:aeff}\end{equation}

These two equation define a deterministic time evolution in ($J,a$)
space from given initial values ($J(0)\!=\! J_{0},\, a(0)\!=\! a_{0}$),
to a final point ($J_{f},a_{f}$), where $s\!=\!\scr$. To recover
$W(J;a_{0})$ at $s_{\mathrm{crit}}$, we introduce some scatter in
the effective velocities $\mathrm{d}J/\mathrm{d}t$,

\begin{equation}
\frac{\mathrm{d}j}{\mathrm{d}t}=-\xi y\frac{j_{m}^{2}(a_{0})}{t_{h}},\label{e:Jeff}\end{equation}
 where $j$ denotes angular momentum in terms of $J_{lc}$, $y$ is
drawn from the positive wing of the normalized Gaussian distribution,
$G_{1}(y)$, and $\xi\!\sim\!1$ (a free parameter) is the width of
the distribution, which can vary depending on the system's parameters.
Equations (\ref{e:aeff},\ref{e:Jeff}) can be solved analytically,

\begin{equation}
j(t)\!=\! j_{0}\!-\! y\xi j_{m}^{2}(a_{0})\frac{t}{t_{h}}\,,\qquad a(t)\!=\!\left[1\!-\!\frac{(d_{c}/a_{0})^{3/2}}{6\xi yj^{6}}\right]^{2}a_{0}\,.\label{e:at}\end{equation}
 The stopping condition $s(j,a)=s_{\textrm{crit}}$ gives an additional
relation between $j$ and $a$, so that the final values are related
by $a_{f}=\scr^{2/3}j_{f}^{-10/3}d_{c}.$ This ties the initial and
final values through the equations of motions

\begin{equation}
y(j_{f})=\frac{(6\xi)^{-1}(d_{c}/a_{0})^{3/2}j_{f}^{-6}}{1-\scr^{1/3}(d_{c}/a_{0})^{1/2}j_{f}^{-5/3}}.\label{e:bJ}\end{equation}

The relation between the initial distribution of $j$-velocities and
the final $j$ distribution at is $W(j_{f};a_{0})=G_{1}(y)[\mathrm{d}y(j_{f})/\mathrm{d}j_{f}]\,.$
The integrated DF over all the cusp stars is

\begin{equation}
W(j_{f})\mathrm{d}j_{f}=\frac{\int^{a_{c}}N(a)W(j_{f};a)\,\mathrm{d}a}{\int^{a_{c}}N(a)\,\mathrm{d}a}\,.\label{e:WJ}\end{equation}

\subsection{Comparison of the different methods}

\label{ss:compare}

We compare the results of the MC simulation, the diffusion-dissipation
equation and an integration of equations (\ref{e:Jeff}) and (\ref{e:aeff}).
In all cases the calculation is stopped when $s\!=\!10^{-3}$; at
that point the eccentricity is still approximately unity, but scattering
becomes entirely negligible so that even in the MC simulation the
quantities evolve essentially deterministically according to equations
(\ref{e:DEgw}-\ref{e:g(e)}). The calculations assume a cusp with
slope $p\!=\!0$, $\Mbh\!=3\!\times\!10^{6}\,\Mo$ and $\Ms\!=\!0.6$
(corresponding to WDs).

\begin{figure}
\plotone{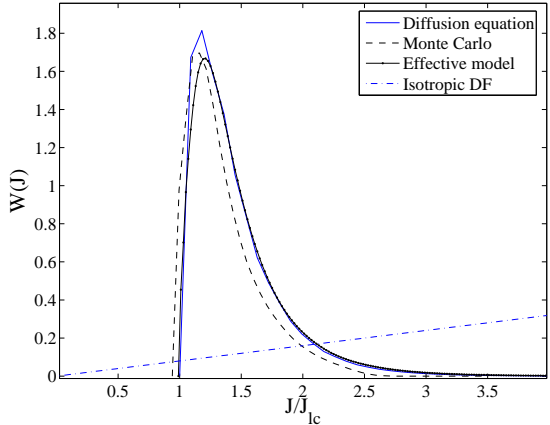}

\figcaption[FileName]{\label{f:Jlossdist}A comparison of the DFs of angular momenta for
inspiraling stars in a $p\!=\!0$ stellar cusp, derived from (1) MC
simulations, (2) direct integration of the diffusion-dissipation equation
(Eq. \ref{e:diffdiss}) and (3) the effective model (Eq. \ref{e:WJ},
with $\xi\!=\!0.6$). The DFs are normalized over the displayed $J$-range.
The DFs are shown at $s\!=\!10^{-3}$, when the dynamics are dominated
by dissipation and scattering no longer plays a role, but the orbits
are still well outside the \LISA\ band, $P\!\gg\! P_{L}$ and $e\!\rightarrow\!1$.
This DF sets the initial conditions for the dissipation phase. The
DF of the isotropic parent population (no sink, no dissipation; Eq.
\ref{e:Niso}) is also shown for contrast. Unlike the DF of the inspiraling
stars, it is dominated by high-$J$ stars.}
\end{figure}

Figure (\ref{f:Jlossdist}) shows the very good correspondence between
the three approaches, whose underlying assumptions and approximations
are summarized in table (\ref{t:comp}). One important conclusion
is that all stars enter the dissipation phase with very small angular
momenta, $1\!\lesssim\! J/J_{lc}\!\lesssim2$. For a suitable choice
of $\xi\!\sim\!1$, the actual complicated stochastic tracks are well
mimicked by the tracks of the {}``effective stars''. The semi-analytical
approach not only identifies the correct scale of angular momentum
of inspiraling stars, but reproduces the DF. Its practical value lies
in the relative ease of its application compared to the time consuming
MC simulations or the integration of the diffusion-dissipation equation. 


\begin{table}

\caption{\label{t:comp}Comparison of assumptions in the 3 methods}

\begin{center}\begin{tabular}{lcccc}
\hline 
Method&
GR potential&
 $\Delta_{1}J_{{\rm scat}}$&
 $\Delta J_{{\rm GW}}$&
 Stop at $s_{{\rm crit}}$\tabularnewline
\hline 
 Monte Carlo&
yes&
 yes &
 yes &
 no$^a$\tabularnewline
 Diffusion/Dissipation&
no&
no&
 no &
 yes\tabularnewline
 Effective track&
no&
no&
 no&
 yes\tabularnewline
\hline
\multicolumn{5}{l}{$^{a}$ For the purpose of comparison to the other two methods,}\tabularnewline
\multicolumn{5}{l}{ the MC simulation was terminated at $s_{\rm crit}$.}
\tabularnewline
\hline
\end{tabular}\end{center}
\end{table}


\section{Inspiral rates and distribution of orbital parameters}

\label{s:results} 

We now proceed to apply the MC simulation to different types of COs
inspiraling into a $3\!\times\!10^{6}\,\Mo$ MBH in a galactic center,
and into a $10^{3}\,\Mo$ IBH in a stellar cluster. We find from the
simulations the critical semi-major axis $a_{c}$ and calculate the
DF of the eccentricities of \LISA\   sources. We stop the simulations
when the orbital period falls below the longest period detectable
by \LISA\  , $P_{L}\!=\!10^{4}\,\mathrm{s}$, which corresponds to
a semi-major axis of 

\begin{equation}
a_{L}=23.5M_{6}^{-2/3}r_{S},\label{e:aL}\end{equation}
where $\Mbh=10^{6}M_{6}\,\Mo$.We record the eccentricity at that
point and construct the DF of the eccentricity at $P_{L}$. It is
straightforward to integrate the orbits in the eccentricity histograms
backward (forward) in time to larger (smaller) values of $P$, see
e.g. Barack \& Cutler (\cite{BC04a}).

The stars are distributed according to a powerlaw distribution
with different values for $p$. The total number of stars within $r_h$
was assumed to be $N_h=2\Mbh/\Mo$, with different number fractions for
the respective species. See table (\ref{t:stars}) for the assumed
parameters of the stellar populations.

\subsection{Massive black holes in galactic centers}

\label{s:MBH}

We assume a $\Mbh\!=\!3\!\times\!10^{6}\,\Mo$ MBH as a representative
example and make the simplifying assumption that the MBH is not
spinning.  Real MBHs probably have non-zero angular momentum (e.g. see
for evidence of spin of the MBH in our Galactic Center Genzel et
al. \cite{GS03}). An important qualitative difference is that in the
Schwarzschild case the eccentricity of a GW-emitting star decreases
monotonically with time, which this is not so for non-zero angular
momentum (e.g. Glampedakis et al. \cite{GHK02}).  We conducted several
MC simulations with Kerr metric orbits and verified that in spite of
changes in details, our overall results hold.

The tidal field of MBHs in this mass range disrupts main sequence
stars before they can become detectable \LISA\ sources. A possible
exception is our own Galactic Center, where the weak GW emission from
very low mass main sequence stars (which are the densest and so the
most robust against tidal disruption) may be detected because of their
proximity (Freitag \cite{FR03}). However, here we only consider GW
from COs. Table (\ref{t:stars}) summarizes the parameters assumed
for the properties of white dwarfs (WDs), neutron stars (NSs) and
stellar mass black holes (BHs). The BHs are assumed to be strongly
segregated in a steep cusp because of their much heavier mass (Bahcall
\& Wolf \cite{BW77}). The total (dark) mass in COs near MBHs is not
known, but in future it may be constrained by deviations from Keplerian
motions of luminous stars very near the MBH in the Galactic center
(Mouawad et al. \cite{MO04}).


\begin{table}

\caption{\label{t:stars} Parameters for stellar populations and inspiral rates}

\begin{center}\begin{tabular}{ccccccccc}
\hline 
Star &
$\Ms$&
 $t_{h}$&
$r_{h}$&
 $p$&
 $f_{s}$&
$a_{c}^{c}$&
$a_{c}^{d}$&
 $\Gamma_{i}$\tabularnewline
&
 {[}$\Mo${]}&
 {[}Gyr{]}&
{[}pc{]}&
&
&
{[}mpc{]}&
{[}mpc{]}&
 {[}$\mathrm{Gyr^{-1}}${]}\tabularnewline
\hline
WD$^{a}$&
 0.6 &
 $10$ &
2&
 0 &
 $0.1$&
30&
20&
 7.8 \tabularnewline
NS$^{a}$&
 1.4 &
$5$&
2&
 0 &
 0.01&
40&
20&
 1.8 \tabularnewline
BH$^{a}$&
 10 &
$1$&
2&
 1/4 &
 $1(-3)$&
20&
10&
 4.7 \tabularnewline
NS$^{b}$&
 1.4 &
$1(-3)$ &
0.05&
 0 &
 0.01&
2&
1&
 4.3 \tabularnewline
BH$^{b}$&
 10 &
$1(-3)$ &
0.05&
 1/4 &
 1(-3)&
5&
2&
 6.7 \tabularnewline
\hline
\multicolumn{9}{l}{$^{a}$ MBH with $\Mbh\!=\!3\times10^{6}\,\Mo$}\tabularnewline
\multicolumn{9}{l}{$^{b}$ IBH with $\Mbh\!=\!10^{3}\,\Mo$.  }\tabularnewline
\multicolumn{9}{l}{$^{c}$From equation (\ref{e:ac})}\tabularnewline
\multicolumn{9}{l}{$^{d}$From the MC simulations}\tabularnewline
\hline
\end{tabular}\end{center}
\end{table}

\begin{figure}

\plotone{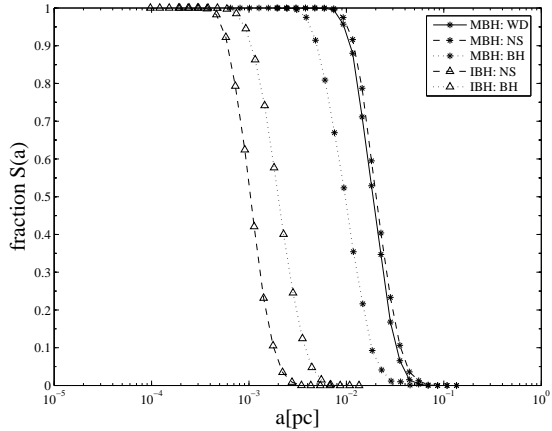}

\figcaption[FileName]{\label{f:GWfrac}Dependence of the normalized fraction of inspiral
events on initial semi-major axis, for the case of a MBH of $\Mbh\!=\!3\times10^{6}\,\Mo$ (asterisks)
and an IBH (triangles). The solid line is for white dwarfs, the dashed
lines for neutron stars, and the dotted lines for stellar mass black
holes. For the parameters of these cases see table (\ref{t:stars}).}
\end{figure}

Fig. (\ref{f:GWfrac}) shows the normalized inspiral probability
function $S_{s}(a_{0})$ , where $s$ stands for WD, NS, and BH. The
critical semi-major axis is similar for WDs and NSs, but much smaller
for BHs, because of the smaller relaxation time and the higher central
concentration due to mass segregation. The functions $S_{s}(a)$ are
used to calculate the inspiral rates (Eq. \ref{e:SU}) that are listed
in table (\ref{t:stars}).  The rate for WDs is highest, $\Gamma_{\rm
WD}=7.8\,{\rm Gyr^{-1}}$. We find that because of mass segregation,
the rate for stellar BHs is of the same order of magnitude as that for
WDs, $\Gamma_{\rm BH}=4.7\,{\rm Gyr^{-1}}$ although their number
fraction is lower by two orders of magnitude.  The hierarchy of rates
in table (\ref{t:stars}) is not, of course, necessarily the same as
cosmic rates that \LISA\ will observe; NSs and BHs are more massive
than WDs, and can be observed at larger distances.

\begin{figure}

\plotone{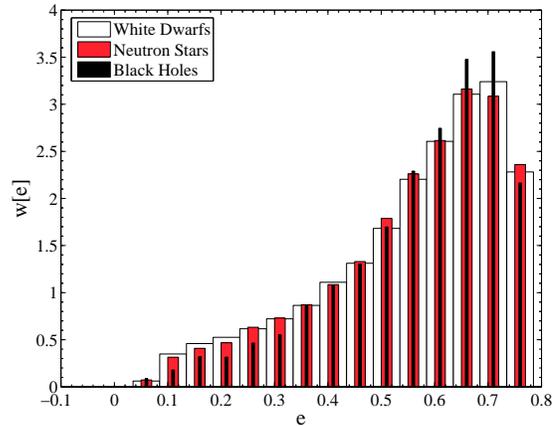}

\figcaption[FileName]{\label{f:WeMBH}The probability DF of a CO entering the \emph{\LISA}
band with eccentricity $e$ for a MBH of mass $\Mbh\!=\!3\!\times\!10^{6}\,\Mo$.
The histograms represent WDs, NSs and BH (from broad to narrow). Note
that the maximal possible eccentricity for which $P\!<\! P_{L}$ is
$e_{\textrm{max}}\!=\!0.81$. See table (\ref{t:stars}) for the cusp
parameters.}
\end{figure}

Fig. (\ref{f:WeMBH}) shows the eccentricity DFs of stars on orbits
with $P\!=\! P_{L}$; note that since $a\!=\! a_{L}$ is fixed, the
orbits are fully determined by $e$. The DFs show a strong bias to
large eccentricities. The maximal eccentricity possible for a star
orbiting a $3\!\times\!10^{6}\,\Mo$ MBH in the \emph{\LISA} band
is $e_{\textrm{max}}\!=\!0.81$ (Eq. \ref{e:eL}). 

It should be emphasized that the histogram in figure (\ref{f:WeMBH})
can be obtained deterministically from the DF $W(J;s_{\mathrm{crit}}\!=\!10^{-3})$
in figure (\ref{f:Jlossdist}), because the stars have already reached
the point where scattering is negligible. This would not be the case
had the MC simulation been terminated at $s_{\mathrm{crit}}=1$ (e.g.
Freitag \cite{FR03}), since then significant scatterings would continue
to redistribute the orbital parameters. We find however that the final
DF (at $P\!=\! P_{L}$), which is obtained by integrating $W(J;s_{\mathrm{crit}}\!=\!1)$
forward in time according to the GW dissipation equations (Eqs. \ref{e:DEgw},\ref{e:DJgw})
without scattering, is not much different from that shown in Fig.
(\ref{f:WeMBH}). This coincidence is due to the fact that the stars
with the largest eccentricities, which typically drop into the MBH,
are replenished by the stars with slightly lower eccentricities. The
main consequence of choosing $\scr$ too large is an overestimate
of the total event rate; stars which actually fall into the MBH are
erroneously counted as contributing to the GW event rate. Incidentally,
even though stars that do not complete the inspiral are not individually
resolvable, they will contribute to the background noise in the \LISA\ band
(Barack \& Cutler \cite{BC04b}). 

A premature neglect of the effects of scattering in previous studies
(e.g. Freitag \cite{FR01}) probably explains in part why our derived
rates are significantly lower. Those studies usually assumed that
stars will spiral in without further perturbations once $s\!=\!1$. We
ran a simulation that was stopped at $s_{\mathrm{crit}}\!=\!1$ instead
of $s_{\mathrm{crit}}\!=\!10^{-3}$. The event rate in that unrealistic
case is about $\sim\!6$ times higher. This does not explain all of the
discrepancies, which are hard to track as different methods are
used. The different stopping criterion may also explain why our rates
are lower than those estimated by Hils \& Bender (\cite{HB95}), who
used a method similar to ours, without specifying the criterion for
inspiral.

\subsection{Intermediate mass black holes in stellar clusters}

Unlike MBHs with masses $\Mbh\!\gtrsim\!10^{6}\,\Mo$ , there is little
firm observational evidence at this time for the existence of IBHs
with masses $\Mbh\!\sim\!10^{3}\,\Mo$ (see review by Miller \& Colbert
\cite{MC04}). However, there are plausible arguments arguing in favor
of their existence. From a theoretical point of view, these objects
are thought to form naturally, such as in population III remnants
(Madau \& Rees \cite{MR01}), or in a runaway merger of young stars
in dense stellar clusters (Portegies Zwart \& McMillan \cite{PZM01};
Portegies Zwart et al. \cite{PZea04}). From an observational perspective,
IBHs may power some of the ultraluminous X-ray sources (e.g. Miller,
Fabian, \& Miller \cite{MFM04}), for example by tidal capture of
a main sequence companion star (Hopman et al. \cite{HPZA04}).

For the purpose of estimating the orbital parameters of GW emitting
stars, we assume that IBHs lie at the center of dense stellar clusters
(see model parameters listed in table \ref{t:stars}). White dwarfs are
disrupted by an IBH before entering the \emph{\LISA} band, so that
only the most compact sources, neutron stars and stellar mass BHs can
emit GW in the \LISA\ frequency band. The same values for the stellar
population fractions $f_s$ where taken here as for MBHs. We note that
this is not necessarily the case. For example, N-body simulations
indicate that a large fraction of BHs may be ejected in an early stage
of the cluster's life if a massive stellar object forms a tight binary
with the IBH (Baumgardt et el. \cite{Baum04b}).

\begin{figure}

\plotone{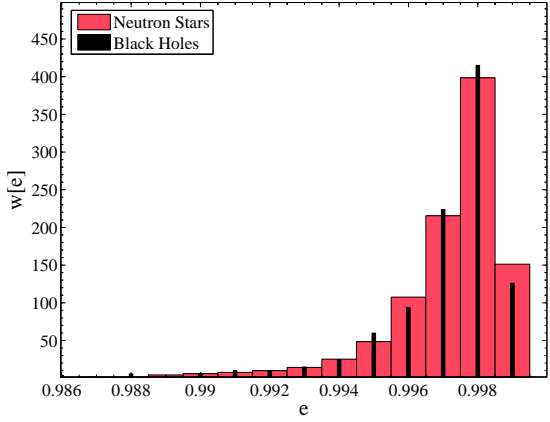}

\figcaption[FileName]{\label{f:IBHecc}The probability DF of a compact remnant entering
the \emph{\LISA} band with eccentricity $e$ for an IBH of mass $\Mbh\!=\!10^{3}\,\Mo$.
Only NSs (broad) and BHs (narrow) are considered; WDs are probably
disrupted by the tidal field. For an IBH, the maximal possible eccentricity
for which $P\!<\! P_{L}$ is nearly unity, $e_{\textrm{max}}\!=\!0.998$,
all inspiraling stars are likely to have eccentricities close to the
maximum value. See table (\ref{t:stars}) for the cusp parameters.}
\end{figure}

Figure (\ref{f:GWfrac}) shows the inspiral probability functions
$S_{s}(a)$, and Fig. (\ref{f:IBHecc}) shows the eccentricity histograms
of stars on orbits with $P=P_{L}$. The maximum eccentricity still
observable by \LISA\ is nearly unity, and the IBH case shows even
more clearly the strong tendency towards large eccentricities. In
general, this effect becomes more prominent for lighter BHs. The high
eccentricity makes the GW signal highly non-monochromatic. The star
spends most time at apo-apse, emitting a relatively weak, low frequency
signal. At periapse short pulses of high frequency GW is emitted.
For IBHs of $\Mbh\!\sim\!10^{3}\,\Mo$, the frequency $\nu_{p}$ of
these short bursts at periapse is of the order $\nu_{p}\lesssim100$
Hz. This is too high to be measurable by \LISA\ , but may be measurable
by ground based detectors such as \textit{LIGO} or \textit{VIRGO}.

As anticipated (Eq. \ref{e:massdependence}), the rate of inspiraling
stars is comparable to that of a MBH (Table \ref{t:stars}). However,
due to the extremely high eccentricities and small semi-major axes
of the orbits, the GW emission is very efficient and they spiral in
on a very short time scale (on the order of a year).

\section{Summary and discussion}

\label{s:sum}

Stars spiraling into MBH due to the emission of GW are an important
potential source for future GW detectors, such as \LISA. The detection
of the signal against the noise will be challenging, and requires
pre-calculated wave-forms.  The waveforms depend on the orbital
parameters of the inspiraling stars. Our main goal in this study was
to derive the distribution of eccentricities of inspiraling stars. The
orbital statistics of such stars reflect a competition between orbital
decay through dissipation by the emission of GW, and scattering, which
deflects stars into eccentric orbits but can also deflect them back to
wider orbits or straight into the MBH.

Inspiral is a slow process, and unless the stars start close enough to
the MBH, they will be scattered off their orbit. We identified a
critical length scale $a_{c}$ which demarcates the volume from which
inspiral is possible. We obtained an analytical expression for the
inspiral rate and showed that it is of the order of a few per Gyr per
galaxy, much smaller than the rate for direct capture (Alexander \&
Hopman \cite{AH03}), that it is nearly independent of the relaxation
time for typical stellar cusps, and that it grows slowly with
decreasing MBH mass (assuming the $\Mbh-\sigma$ relation). Throughout
we assumed a single powerlaw DF. Generalization to different profiles
is straightforward. Qualitatively, the inspiral rate is determined by
the stellar density near $a_c$, and is not very sensitive to the exact
profile far away at radii much smaller or much larger than $a_c$. The
rate of GW events depends on the number of COs inside $a_{c}$ (which
is much smaller than the MBH radius of influence), and so
mass-segregation can play an important role in enhancing the event
rate by leading to a centrally concentrated distribution of COs.

We obtained a relatively low rate. One important reason for this is
our stringent criterion for inspiral. This may also explain why our
results deviate from Hils \& Bender \cite{HB95}, although they do not
specify their precise criterion for inspiral. Comparison with other
works are more complicated. Possible differences may stem from
different normalizations of the central density, different CO
fractions, different criteria for inspiral, and the way mass
segregation is treated. An essential step in the future will be to
analyze inspiral processes by N-body simulations, which are feasible
already for small systems (Baumgardt et al. \cite{Baum04a},
\cite{Baum04b}; Preto et al. \cite{P04}).

The detection rate depends on the inspiral rates, but also on the mass,
relaxation time, the orbital parameters (especially the eccentricity)
and the details of the detection algorithm (Barack \& Cutler
\cite{BC04a}). Here we provide a simple recipe to estimate the number of
detectable sources.  We assume that the various dependencies above can
be expressed by an effective strain $\hat{h}$.

The strain of GW resulting from a star of mass $\Ms=m\Mo$ orbiting a
MBH of mass $\Mbh=10^6\Mo M_6\gg\Ms$ at a distance $d={\rm
Gpc}\,d_{\rm Gpc}$, on a circular orbit of orbital frequency $\nu=
10^{-3}{\rm s^{-1}}\nu_{-3}$, is given by

\begin{equation}\label{eq:strain}
h = 1.7\times 10^{-23}\nu_{-3}^{2/3}d_{\rm Gpc}^{-1}M_6^{2/3}m.
\end{equation}
(e.g., Sigurdsson \& Rees \cite{SR97}).

The cosmic rate of {\it LISA} events is given by

\begin{equation}\label{eq:cosmrate}
\Gamma_{\rm tot} = 
\int_{M_{\rm min}}^{M_{\rm max}} d\Mbh {d\nbh\over d\Mbh} \Gamma_i(\Mbh) V(\Mbh),
\end{equation}
where $d\nbh/d\Mbh$ is the number of MBHs per unit mass, per unit
volume, and $ V(\Mbh)$ is the volume in which stars can be observed by
{\it LISA}. Aller \& Richstone (\cite{AR02}) used the $\Mbh-\sigma$
relation to ``weigh'' the MBHs. The spectrum is roughly approximated
by

\begin{equation}
{d\nbh\over d\Mbh} = 10^7\left({1\over10^6\Mo}\right) M_6^{-1}\, {\rm Gpc^{-3}}.
\end{equation}

The LISA sensitivity curve at $\nu=10^{-3}$ is $h\sim10^{-23}$ for one
year of observation with signal to noise ratio S/N=1 (see e.g.  {\it
http://www.srl.caltech.edu/lisa}). We adopt this value as a
representative detection threshold for $\hat{h}$. If the effective
strain has to be be at least $10^{-23}\hat{h}_{-23}$ for the star to
be observable, then, for a Euclidean Universe,

\begin{equation}
V(\Mbh) = 20.6\,{\rm Gpc^{3}}\,\hat{h}_{-23}^{-3}\nu_{-3}^2m^3M_6^2.
\end{equation}

Finally, the rate per MBH is 
\begin{equation}
\Gamma_i(\Mbh) =
\Gamma_i(3\times10^6\Mo)(3M_6)^{-1/4},
\end{equation}
where we used the expression for the dependence of the inspiral rate
on the MBH mass, equation (\ref{eq:massdependence}), with $\beta=4$.
The expression can be calibrated with the MC results for a $3\times
10^6 \Mo $ MBH.

Integrating (\ref{eq:cosmrate}) from $M_6 = 0.5$ to  $M_6=5$ gives

\begin{equation}
\Gamma_{\rm tot} = 
1.5\,\peryr\,\left[{\Gamma_i(3\times10^6\Mo)\over {\rm Gyr^{-1}}}\right]m^3\hat{h}_{-23}^{-3}\nu_{-3}^2.
\end{equation}

The number of sources which {\it LISA} can observe at any moment is
estimated by $\mathcal{N} = \Gamma_{\rm tot}\times t_L(\bar{e}, a_L)$,
where $t_L$ is the time a star with eccentricity $\bar{e}$ spends in
the {\it LISA} band before being swallowed; here $\bar{e}$ is the
average eccentricity of stars entering the {\it LISA} band.

For example, our calculations for WD inspiral indicate that
$\Gamma_i(3\times10^6\Mo)=7.8\,{\rm Gyr^{-1}}$. For a circular orbit
with a period of $P=10^3\,{\rm s}$ and $M_6=1$, the inspiral time is
$t_L=54\,{\rm yr}$, in which case the number of detectable sources
would be $\mathcal{N} \sim\,130\hat{h}_{-23}^{-3}\nu_{-3}^2$.

We would like to emphasize that this estimate is to be treated with
caution. The number of detectable sources depends strongly on the
assumptions. In particular, GWs from eccentric sources are not
monochromatic and may be harder to detect. The analysis in this paper
can be used for a more detailed analysis of the number of detectable
{\it LISA} sources.

We used three complementary methods to model the inspiral process, MC
simulations, numerical solution of the diffusion/dissipation equation
and an analytic effective orbit model. We followed the evolution of
the orbital properties of the inspiraling stars to the point where
they decoupled from the scattering. All three methods were in
excellent agreement. We find that the distribution of orbital angular
momenta is strongly peaked near $J\!\gtrsim\! J_{lc}$, with the
detailed form of the distribution depending somewhat on the parameters
of the stellar system. We demonstrated that estimates that
{}``freeze'' the scattering prematurely may lead to erroneously high
rates by counting stars that actually plunge into the MBH. We then
used the most versatile method, the MC simulations, to continue
evolving the orbits in a GR Schwarzschild potential, taking into
account energy and angular momentum loss due to GW emission (and
residual perturbations by scattering).  We derived the distribution of
eccentricities of the inspiraling stars as they enter the detection
window (orbital period of $P_{L}\!\lesssim\!10^{4}\,\mathrm{s}$ for
\LISA).

Our main result is that the eccentricities of stars entering the \LISA\ 
band are strongly skewed toward high values.

\acknowledgements{TA is supported by ISF grant 295/02-1, Minerva grant 8484, and a
New Faculty grant by Sir H. Djangoly, CBE, of London, UK.}

\end{document}